\documentclass[12pt,preprint]{aastex}

\def\etal{{et~al. \,}}

\def\teff{T$_{\rm eff}\,$}

\def\Dwa{$\,$\uppercase\expandafter{\romannumeral5}$\,$}

\def\sles{\lower2pt\hbox{$\buildrel {\scriptstyle <}
   \over {\scriptstyle\sim}$}}

\def\sgreat{\lower2pt\hbox{$\buildrel {\scriptstyle >}
   \over {\scriptstyle\sim}$}}
\def\sharpnull#1{}
\def\aa{Astron. Astrophys.\ }

\begin{document}

\title{Anisotropies in the Neutrino Fluxes and Heating Profiles in
Two-dimensional, Time-dependent, Multi-group  
Radiation Hydrodynamics Simulations of Rotating 
Core-Collapse Supernovae}

\author{R. Walder\altaffilmark{1},  
A. Burrows\altaffilmark{1}, 
C.D. Ott\altaffilmark{2},
E. Livne\altaffilmark{3},
I. Lichtenstadt\altaffilmark{3},
M. Jarrah\altaffilmark{4}}
\altaffiltext{1}{Department of Astronomy and Steward Observatory, 
                 The University of Arizona, Tucson, AZ \ 85721;
                 rwalder@as.arizona.edu,burrows@as.arizona.edu}
\altaffiltext{2}{Max-Planck-Institut f\"{u}r Gravitationsphysik,
Albert-Einstein-Institut, Golm/Potsdam, Germany; cott@aei.mpg.de}
\altaffiltext{3}{Racah Institute of Physics, The Hebrew University,
Jerusalem, Israel; eli@frodo.fiz.huji.ac.il,itamar@saba.fiz.huji.ac.il}
\altaffiltext{4}{Department of Electrical and Computer Engineering, 
                 The University of Arizona, Tucson, AZ \ 85721;
                 mjarrah@ece.arizona.edu}
\begin{abstract}

Using the 2D multi-group, flux-limited diffusion version of the code VULCAN/2D, 
that also incorporates rotation, we have calculated the 
collapse, bounce, shock formation, and early post-bounce evolutionary
phases of a core-collapse supernova for a variety of initial rotation rates.  This is the first
series of such multi-group calculations undertaken in supernova theory with fully multi-D tools.
We find that though rotation generates pole-to-equator angular anisotropies in the neutrino
radiation fields, the magnitude of the asymmetries is not as large as
previously estimated.  The finite width of the neutrino decoupling surfaces and the
significant emissivity above the $\tau=2/3$ surface moderate the angular contrast.
Moreover, we find that the radiation field is always more 
spherically symmetric than the matter distribution, with its plumes and convective eddies.
The radiation field at a point is an integral over many sources from the different contributing directions.
As such, its distribution is much smoother than that of the matter 
and has very little power at high spatial frequencies. 
We present the dependence of the angular anisotropy of the neutrino fields
on neutrino species, neutrino energy, and initial rotation
rate.  Only for our most rapidly rotating model do
we start to see qualitatively different hydrodynamics, but for the lower rates consistent
with the pre-collapse rotational profiles derived in the literature the anisotropies,
though interesting, are modest.  This does not mean that rotation does not play
a key role in supernova dynamics.  The decrease in the effective gravity due to the centripetal
effect can be quite important.  Rather, it means that when a realistic mapping between initial and final
rotational profiles and 2D multi-group radiation-hydrodynamics are incorporated into collapse simulations
the anisotropy of the radiation fields may be only a secondary, not a pivotal factor,
in the supernova mechanism.

\end{abstract}

\keywords{supernovae, rotation, multi-dimensional radiation hydrodynamics, transport, neutrinos}

\section{Introduction}
\label{intro}

The prompt hydrodynamic bounce in core collapse never leads to direct supernova 
explosions in either 1D, 2D, or 3D; neutrino losses and photodissociation
by the shock debilitate it, even for the lowest mass progenitors and accretion-induced
collapse (AIC).  In the Chandrasekhar context, there is just too much mass
between the place the shock originates ($\sim$0.6 M$_{\odot}$)
and the outer boundary ($\ge$1.2 M$_{\odot}$) and
the shock stalls into an accretion shock.  Furthermore,
in spherical symmetry (1D), it has been shown using Boltzmann neutrino
transfer and the best physics that the delayed neutrino mechanism after shock stagnation
does not work \citep{rampp2000,mezz2001,lieben2001,Tod1}.
In 1D, the bounce shock stalls and is not revived, though an increase of only $\sim$25\%
in neutrino heating would lead to explosion.  Such an increase could
arise from as-yet-unknown neutrino effects or overturning motions
in the inner core that could boost the neutrino luminosity, though
neither of these classes of effects has been demonstrated. 

However, in 2D, but using gray neutrino transfer,
numerous simulations result in explosions \citep{herant,bhf,fryer2002},
although they are sometimes weak.  These calculations
demonstrate that neutrino-driven convection in the
so-called ``gain" region near the shock \citep{bethe}
increases the efficiency of neutrino energy deposition,
increases the size of the gain region, and facilitates explosion.
The 2D multi-group calculations of Janka, Buras, and Rampp (2003)
and Buras et al. (2003) employed multiple 1D radial Boltzmann solves 
in lieu of 2D transport.  They obtained a marginal explosion, but did retain
the velocity-dependent terms in the comoving transport equations.
Recent 3D calculations (Janka et al. 2004), using 1D gray transport along radial rays,
reveal that 3D may be marginally better than 2D.  The energy of this 3D explosion
depended on the authors' choice of inner boundary condition.
Nevertheless, multi-D effects seem to be crucial to the 
mechanism of core-collapse supernovae. 

In the last few years, Shimizu et al. (2001), Kotake, Yamada, \& Sato (2003), 
and Madokoro, Shimizu, \& Motizuki (2004) have suggested that rotation
enhances the neutrino flux ($F_{\nu}$) along the rotational axis and increases the pole-equator
contrast in the neutrino heating rate, thereby facilitating a polar explosion.
Shimizu et al. (2001) performed a set of 2D hydrodynamic simulations on a post-bounce structure  
in which neutrino heating, the pole-to-equator flux contrast, and the 
effective temperature of the neutrinosphere were varied.  This parameter
study did not involve transport and did not consistently determine the 
mapping between the rotation rate and the pole-to-equator flux contrast.
Kotake, Yamada, and Sato (2003) used the code ZEUS-2D with a neutrino leakage scheme
to estimate the asphericity in the neutrino flux for 2D rotating collapse models. 
They assumed that the neutrinos are emitted isotropically from 
a spheroidal neutrinosphere and that the neutrino energy density 
($\varepsilon_{\nu}$) and flux exterior to the neutrinosphere are 
related by $F_{\nu}$ = $c \varepsilon_{\nu}$, where $c$ is the speed of light.  Using shellular and cylindrical
initial rotation laws (Ott et al. 2004), their initial core angular velocities
ranged from 2.7 rad s$^{-1}$ to 112 rad s$^{-1}$ (the latter highly unrealistic) and their initial $T/|W|$ ranged
from 0.25\% to 1.5\%.  Kotake, Yamada, \& Sato's most rapidly rotating models achieved final 
$T/|W|$s (where $T$ is the rotational kinetic energy and $W$ is the gravitational energy)
as high as $\sim$14\% and resulted, according to their prescriptions, 
in pole-to-equator contrasts in the neutrino heating rate of from a few to more than ten.
Specifically, in Kotake, Yamada, \& Sato (2003) representative values of these 
contrasts were 1.26, 2.12, 3.84, 5.61, 8.63, 15.2, and 25.9.
Madokoro, Shimizu, \& Motizuki (2004) assumed a pole-to-equator ratio in 
the neutrino fluxes given by the formula $1 + c_2\cos^2(\theta)$ and explored
parametrically the consequences of various degrees of the prolateness and oblateness of the 
neutrinosphere on the explosion of rotating cores.

None of these rotational anisotropy studies involved consistent radiative transfer calculations, 
nor did they derive in the context of a full radiation-hydrodynamic study
the actual connection between rotation and the anisotropy of the neutrino 
radiation field. None of these studies was multi-group, nor were the various
neutrino species (in particular, $\nu_{e}$ and $\bar{\nu}_e$) 
distinguished.  The rotational simulations of Fryer \& 
Heger (2000, 2D) and Fryer \& Warren (2004, 3D) were some of the most
complete explorations to date of the effects of rotation on the supernova mechanism.
However, these authors used a gray, flux-limited, diffusion approach and did not publish
anything concerning the derived anisotropies of the radiation field, nor
on the potential role of such anisotropies in the explosion.  Their focus was 
on the evolution of the angular momentum and on its interaction with the neutrino-driven convective
motions.  Finally, the work of Janka, Buras, \& Rampp (2003) and Buras et 
al. (2003) included in their set of simulations a slowly rotating model 
and was multi-group, but the transport was along radial rays and the pole-to-equator
differences in the heating rates and fluxes could not be reliably determined.

In this paper, we present new results using the 2D multi-group, flux-limited diffusion (MGFLD)
variant of the multi-group, multi-angle, time-dependent radiation-hydrodynamics code
VULCAN/2D\footnote{VULCAN/2D is a multi-group, multi-angle, time-dependent 
radiation-hydrodynamics code. In addition to being 6-dimensional
(1(time) + 2(space) + 2(angles) + 1(energy groups)),
it has an ALE (Arbitrary
Lagrangian-Eulerian) structure with remap, is axially-symmetric,
can handle rotation, is flux-conservative, smoothly matches to the diffusion limit,
and is implicit in its Boltzmann solver.  However, it does not yet have all the
velocity-dependent terms in the transport equation, such as the Doppler shift
and aberration, though it does have the advection
term.  In addition, it does not currently have energy
redistribution due to inelastic scattering.} (Livne et al. 2004)
to explore the effect of rotation on the anisotropy of the neutrino
radiation field and to test the flux and heating asymmetry estimates 
of Shimizu et al. (2001), Kotake, Yamada, \& Sato (2003),
and Madokoro, Shimizu, \& Motizuki (2004).  Ours are 
the first consistent time-dependent calculations of this
effect. The 2D MGFLD version of VULCAN/2D is computationally much
faster and allows us to more quickly explore 
the model space.  In 2D, we simulate  
collapse, bounce, neutrino shock breakout,
and the neutrino-driven convection stages for
a 11 M$_{\odot}$ progenitor \citep{woosley} both without and with rotation.
Five models with rotation are simulated (\S\ref{aniso}).
Eight neutrino energy groups and three neutrino species ($\nu_e$, $\bar{\nu}_e$, and
``$\nu_{\mu}$") are followed.  The code is parallelized
in energy groups using MPI.  The fluxes are vector fluxes in two dimensions
and both the regions at high and low optical depths are seamlessly followed. 
Rather than imposing a fixed luminosity inner boundary
condition or cutting out the inner core (Scheck et al. 2004), 
the full time-dependence of the emerging neutrino luminosities
and spectra and the motion of the inner core are consistently 
obtained from the simulations. 

The calculations of this paper were performed using the MGFLD variant of VULCAN/2D, and not its full
angle-dependent Boltzmann transport variant.  Runs using the latter require significantly more
CPU-hours to complete (Livne et al. 2004).  Therefore, that variant is well less suited to the parameter study we
present here.  Differences between the MGFLD and Boltzmann versions are small at high
and intermediate optical depths, but can be larger in the semi-transparent to transparent regime.
In particular, at radii of $\ge$200 km, the MGFLD version, by dint of its
diffusive nature, tends to smooth the angular dependence of the radiation
fields and smear the contrast in the neutrino energy densities.  
This limitation should be borne in mind.  However, the
full transport version, using as it does the S$_n$ method to resolve
angular space, can itself introduce anomalies (periodicities) in the radiation fields at radii
larger than $\sim$300-400 km, depending upon the number of angles used to cover the hemisphere.
Hence, all approaches have their numerical limitations, though our qualitative conclusions
concerning the flux and net gain anisotropies interior to the shock wave and
in the inner regions around the neutrinospheres remain robust.  In these regions, the 2D MGFLD
approach provides a very reasonable representation of the multi-species, multi-group neutrino fields.

The limiter we employ is a 2D vector generalization of Bruenn's scalar flux limiter (Bruenn 1985).
It functional form is:

\begin{equation}
\Lambda = \frac{3}{3 + |R|},
\end{equation}
where $R = \frac{\lambda_{\nu} d\ln E_{\nu}}{ds}$, $E_{\nu}$ 
is the neutrino energy density spectrum, and $\lambda_{\nu}$ is the 
total mean-free path. $|d\vec{s}|$ is the differential magnitude 
of the vector distance along the flux direction,
given, as is assumed in the diffusion approximation, by the direction of the gradient of $E_{\nu}$.
Burrows \etal (2000) compared the accuracy with which various 1D flux limiters reproduced
the neutrino heating rates behind the shock wave in the gain 
region found by the SESAME spherical Boltzmann code\cite{B2,Tod1}
and found that this simple limiter performed rather better than others in the literature.

We find that, while rotation does 
indeed induce an anisotropy in the neutrino flux, 
this effect is not necessarily rotation's dominant
consequence.  It is also not 
as large as previously estimated.  We do find 
that for rapid rotation the neutrino heating rate and the 
entropy due to neutrino heating are larger along the poles 
than the equator.  However, other effects of rotation, in 
particular the consequent decrease in the effective gravity 
or the rotation-induced anisotropy in the mass accretion 
rate, may contribute as well (Burrows, Ott, \& Meakin 2003;
Burrows et al. 2004; Yamasaki \& Yamada 2004) and 
be more important in facilitating a robust explosion.  This 
possibility will be the subject of a subsequent paper 
(Burrows et al. 2005)\footnote{Burrows et al. (2004) 
and Burrows, Ott, \& Meakin (2003) have
suggested that the bipolar structures seen in Cas A \citep{will02,will03,hwang04}
and inferred from the polarization of Type Ic supernovae \citep{wang1,wang2}
are a consequence of the neutrino mechanism in the context of rapid rotation,
naturally producing 30$^{\circ}$-60$^{\circ}$ wide-angle ``jets."}.

In \S\ref{aniso}, we discuss the initial rotational profiles we employ
and our initial models.   In \S\ref{basic}, the overall hydrodynamic 
behavior of the rotating and non-rotating models through the first
175 milliseconds (ms) after bounce are described.  The entropy, density ($\rho$),
and velocity evolution during this phase are provided.  
We present the mapping between the initial and ``final" rotational
profiles.  The latter have been consistently derived in the context
of multi-D radiation-hydrodynamic simulations and give one a
reliable estimate of the spin rates expected, given the initial models assumed,
of the protoneutron-star/protopulsar at this epoch in its early evolution. In \S\ref{fluxaniso}, 
we present our major results concerning the angular distributions 
of the neutrino energy fluxes, the corresponding neutrino energy densities,
and the neutrino-matter heating rates during the early post-bounce stages.
In \S\ref{conclusion}, we summarize our general conclusions
concerning the rotation-induced anisotropy of the neutrino radiation
fields in stellar collapse.

\section{The Initial Models Used to Study the Rotation-Induced 
Anisotropy of the Neutrino Field and Heating Profile}
\label{aniso}

For our simulations with rotation we take a non-rotating 11 M$_{\odot}$ 
progenitor model from Woosley \& Weaver (1995) and impose a rotation law (Ott et al. 2004):

\begin{equation}
\label{eq:rotlaw}
\Omega(r) = \Omega_0 \, \bigg[ 1 + \bigg(\frac{r}{{\rm A}}\bigg)^2 \bigg]^{-1}\, ,
\end{equation}
where $\Omega(r)$ is the angular velocity, $r$ is the distance from
the rotation axis, and $\Omega_0$ and A are free parameters that
determine the rotational speed/energy of the model and the
distribution of angular momentum.  Equation (\ref{eq:rotlaw}) starts
the matter rotating on cylinders.

The five rotating models of this study
have $\Omega_0 = 2.68$ rad s$^{-1}$ (Model A), $\Omega_0 = 1.34$ rad s$^{-1}$ (Model B),
$\Omega_0 = 0.6$ rad s$^{-1}$ (Model C), $\Omega_0 = 0.15$ rad s$^{-1}$ (Model D), and
$\Omega_0 = 0.04$ rad s$^{-1}$ (Model E), all with ${\rm A} = 1000$ kilometers.
The corresponding initial $T/|W|$s are 0.29\%, 0.075\%, 1.5$\times 10^{-2}$\%, 
9.4$\times 10^{-4}$\%, and 6.7$\times 10^{-5}$\%, respectively,
and the corresponding initial central periods are 2.34, 4.69, 10.47, 
41.89, and 157.1 seconds, respectively.  
A listing of the six models of this paper and their initial rotational characteristics
is given in Table \ref{table:models}.  Note that using the rotation law of 
eq. (\ref{eq:rotlaw}) makes the inner core rotate much more quickly
than the periphery and puts much of the rotational kinetic energy
in the interior.  The material exterior to A (in this case, 1000 km) is
rotating much more slowly.  For instance, at 2000 km, the spin period 
for Model A is 11.7 seconds and that for Model B is 23.45 seconds.
This material is accreted through the stalled shock wave within 
the first 100's of milliseconds of bounce.

Many of the rotating progenitor models in the recent literature
with comparable rotation rates were calculated without the centrifugal
term turned on, either after the onset of core carbon burning
or at all \citep{heger00,langer,heger04}.  The result is that many of these models initially
expand when mapped into 2D rotation codes such as VULCAN/2D.  Given this, and
the fact that the most current rotating progenitor models were available
only after we began our calculations (which take approximately
one month to complete), we have deferred the study of the rotating
models now in the literature to a later date.  Furthermore, the two main
groups \citep{heger00,langer,hirschi,heger04,meynet} performing
these detailed progenitor simulations up to the onset of collapse
do not agree even qualitatively on the proper prescriptions
for the transport of angular momentum during the various
nuclear burning stages.  The major sticking point is the role of
magnetic fields and their proper treatment.  As a result, there
is still a spread by factors of from 10 to 100 in the specific core angular
momenta and spin periods for the same progenitor ZAMS mass and initial
surface velocity, with magnetic models resulting in
the slowest spin rates.  For instance, Heger, Woosley, \& Spruit (2004),
using different presciptions for angular momentum transport, with and without
magnetic fields, derive core $\Omega_0$s that range from  
0.1 rad s$^{-1}$ to $\sim$6.0 rad s$^{-1}$, with a preference for the
lower values.  However, Hirschi, Meynet, \& Maeder (2004) derive $\Omega_0$s 
near 1.0 rad s$^{-1}$.  Clearly, it will be important to resolve
these issues, since rotation is certainly a
factor in core-collapse phenomenology.

For the non-rotating control model, we employ the non-rotating
11 M$_{\odot}$ progenitor from Woosley \& Weaver (1995) from which we generated
the above rotating Models A-E, and designate this non-rotating Model F.
The grid we use for all models is similar to that described and plotted in Ott et al. (2004), 
but with 81 angular and 128 radial bins. The inner 20 kilometers is tiled more densely 
to better resolve core bounce.  The cylindrical coordinate system allows
the central core, always handled in 2D, to move along the axis of symmetry
if the dynamics requires it.  We use eight energy groups 
centered at 2.5, 6.9, 12, 21, 36.7, 64, 112, and 196.5 MeV.

\section{The Basic Hydrodynamic Behavior}
\label{basic}

Since our focus in this paper is on the magnitude and character
of the rotation-induced anisotropy of the neutrino radiation fields
in the context of core collapse, we will describe only briefly
the hydrodynamic effects of rotation themselves.   More detailed
discussions of the resulting dynamics and of the supernova phenomenon 
as determined in this series of 2D MGFLD simulations 
will be deferred to another paper (Burrows et al. 2005).

\begin{figure}
\figurenum{1}
\centerline{\includegraphics[width=18.cm]{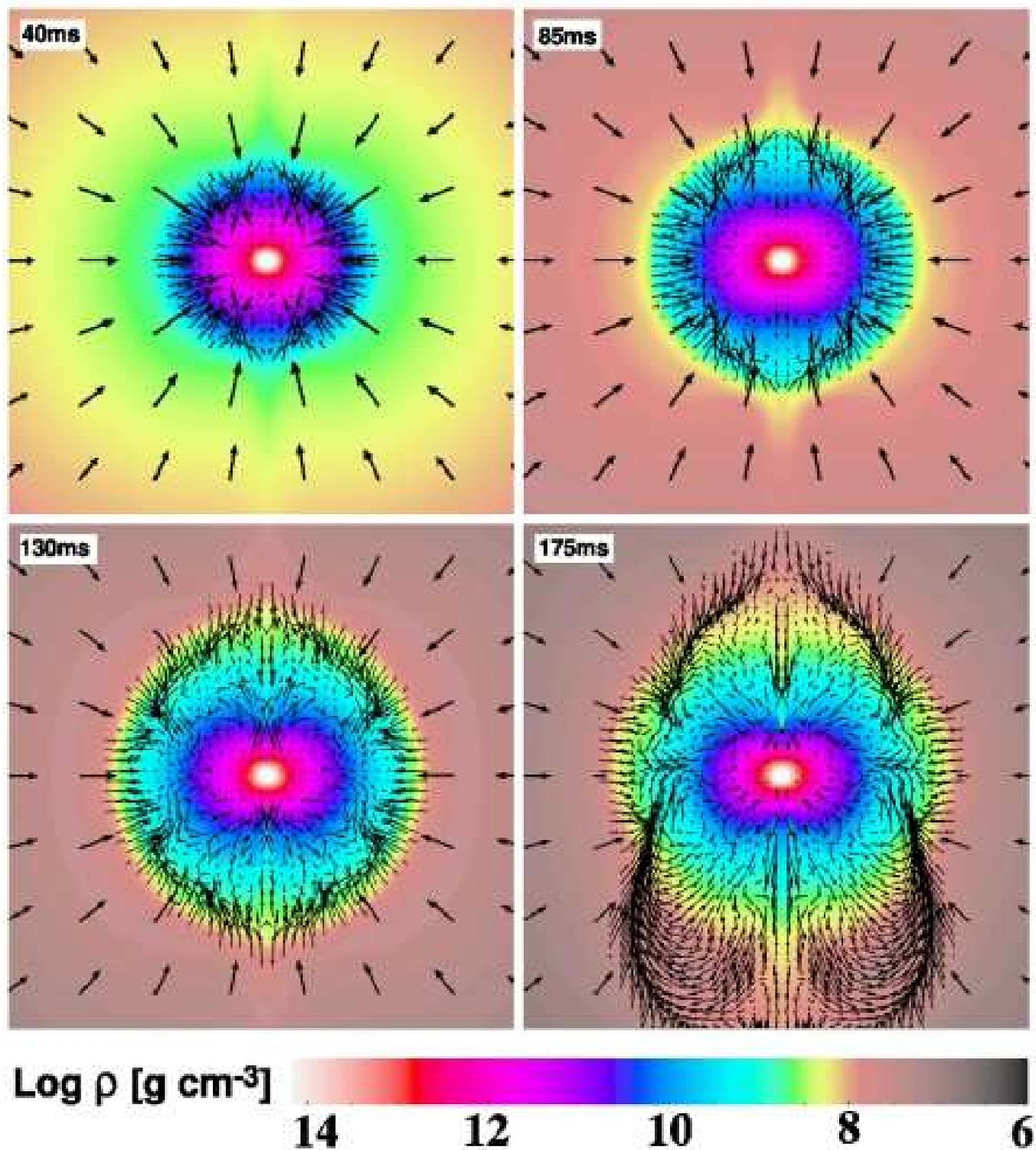} }
\caption{Color map of the mass density ($\rho$) distribution of our fastest rotating Model A
         ($\Omega = 2.68$ rad s$^{-1}$), with the r-z plane projected
         velocities (arrows) superposed, at times 40~ms (upper left panel), 85~ms
         (upper right panel), 130~ms (lower left panel), and 175~ms
         (lower right panel) after bounce. The inner 600~km on a side
         are shown. Note that the
         the arrows which represent the infalling matter are on a sclae 2.5
         times smaller than the arrows representing the shocked
         matter. The inner core is strongly oblate; at 30~km the axis
         ratio of the iso-density contours is approximately 2:1. The
         top-bottom asymmetry develops only late in time and 
         hardly effects the inner high-density region of the 
         protoneutron star.  Note the asymmetry in the position of the shock-wave. 
         Also, compare to the density maps of the slower rotating models shown in
         Fig.~\ref{fig:density.models}.}
\label{fig:density.omega.2.68.times}
\end{figure}

Figure \ref{fig:density.omega.2.68.times} depicts snapshots of 
mass density color maps in the inner 600 
kilometers (km) on a side of Model A ($\Omega_0$ = 2.68 
rad s$^{-1}$) at various times after bounce up to 175 milliseconds.  Velocity vectors
are superposed to trace the flow.  A salient characteristic of this 
plot is the degree of oblateness induced by rotation.  In the inner
$\sim$30 km, the axis ratio of isodensity contours is approximately 2:1, but further out it is
more moderate.  At a given radius of 90 km, the equator-to-pole ratio
in the mass density gradually increases with time starting 
at a value of $\sim$2.5 at 30 ms.  This increase is a 
consequence of the relative decrease in the accretion rate along
the poles due to the centrifugal barrier of rotation 
and the establishment of a funnel (Burrows, Ott, \& Meakin 2003).  
The mapping between the oblateness of the density field 
near the ``neutrinosphere" and the neutrino flux field has been 
a feature in past discussions of the latter's rotationally-induced
anisotropy (e.g., Shimizu et al. 2001; Kotake, Yamada, \& Sato 2003; Janka \& M\"onchmeyer 1989ab).
Note that the position of the shock wave (identified by the clear
jump in color and velocity field) grows steadily with time and that
a top-bottom asymmetry develops at the latest time.  Since this
late-time dipole-like structure is most manifest near the shock wave, and 
the shock at this time is near a radius of $\sim$300 km, 
the top-bottom matter asymmetry does not much affect the
neutrino emissions emerging from the inner 50-100 km.

\begin{figure}
\figurenum{2}
\centerline{\includegraphics[width=18.cm]{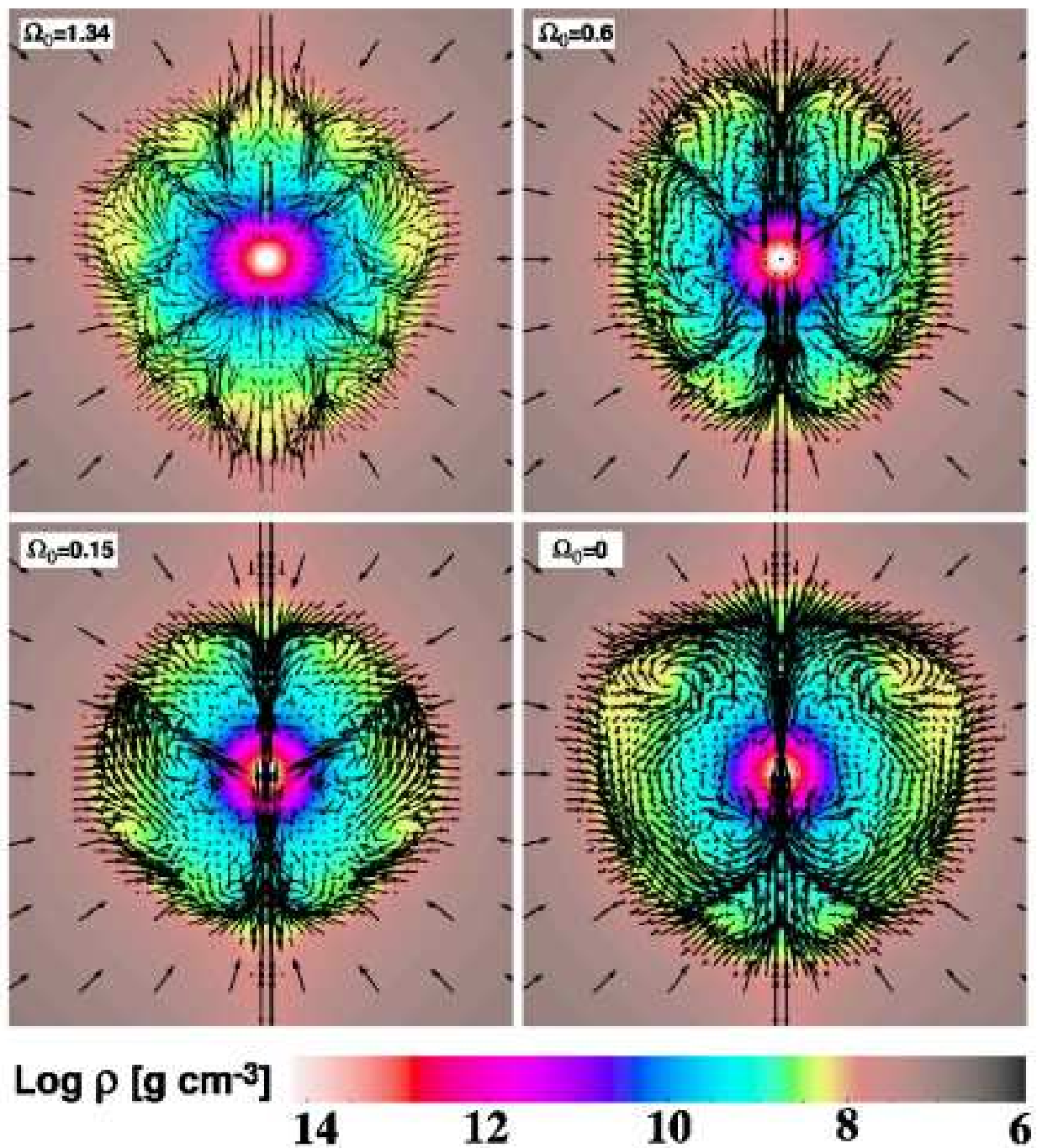} }
\caption{Same as Fig. \ref{fig:density.omega.2.68.times}, but at 175 milliseconds after
         bounce and for Models~B ($\Omega = 1.34$ rad
         s$^{-1}$, upper left panel), C ($\Omega = 0.6$ rad s$^{-1}$,
         upper right panel), D ($\Omega = 0.15$ rad s$^{-1}$, lower
         left panel), and F (non-rotating, lower right panel).  
         Shown is the inner 600~km on a side. In
         comparison to the fastest rotating model A ($\Omega = 2.68$
         rad s$^{-1}$) shown in Fig.~\ref{fig:density.omega.2.68.times}, the density
         distribution is much less prolate. Only Model~B ($\Omega =
         1.34$ rad s$^{-1}$) shows a significant prolateness, whereas
         the models with less rotation than $\Omega = 0.6$ rad
         s$^{-1}$ exhibit close to no rotational flattening. 
         Note the increase in the number of convection cells with
         increasing rotation, from one large-scale convection cell in
         the non-rotating model to five cells in Model~B. However, the
         fastest rotating Model~A manifests only one
         large-scale cell and only some smaller plumes in the
         equatorial region.}
\label{fig:density.models}
\end{figure}

\begin{figure}
\figurenum{3}
\centerline{\includegraphics[width=14.cm]{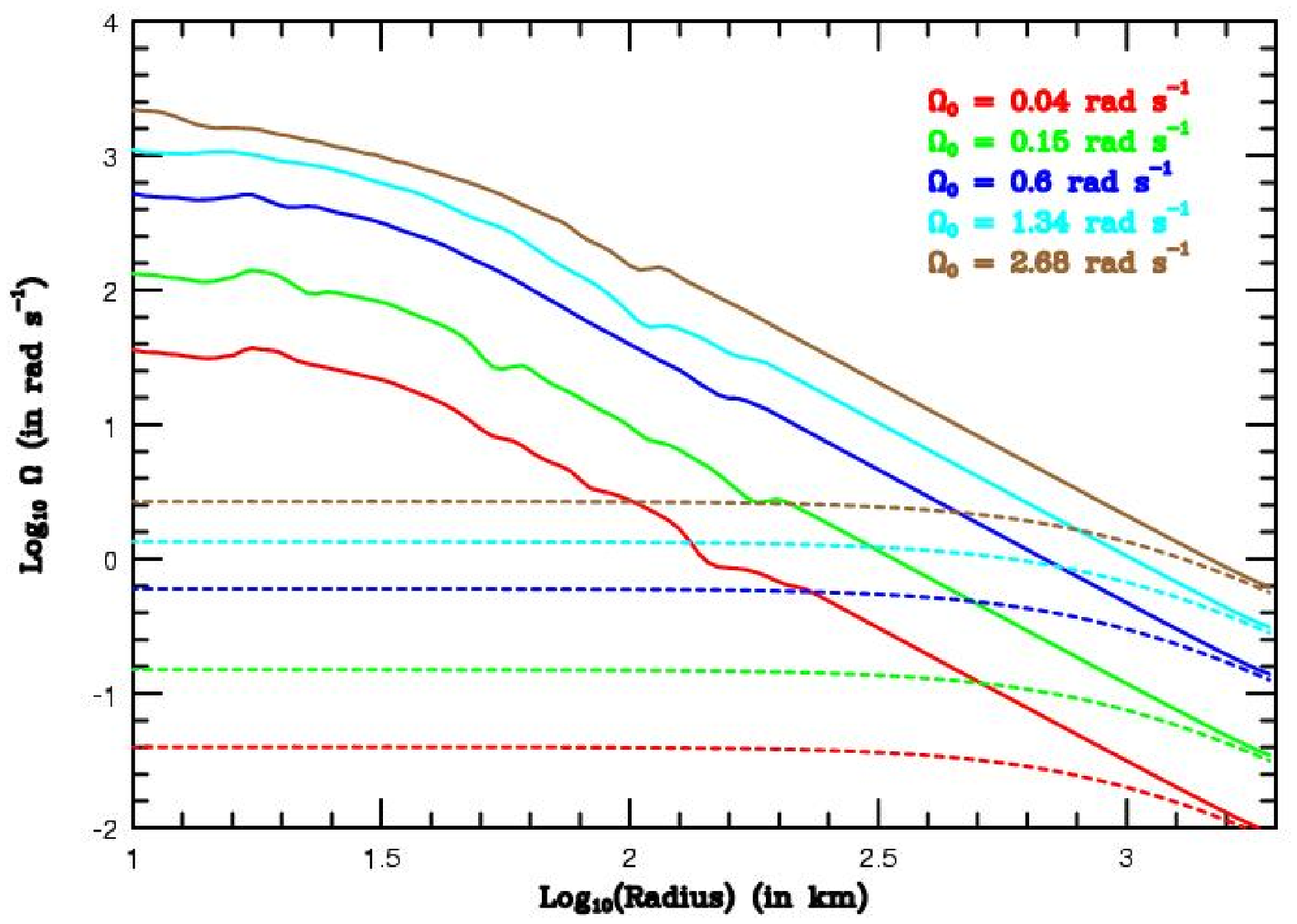}}
\centerline{\includegraphics[width=14.cm]{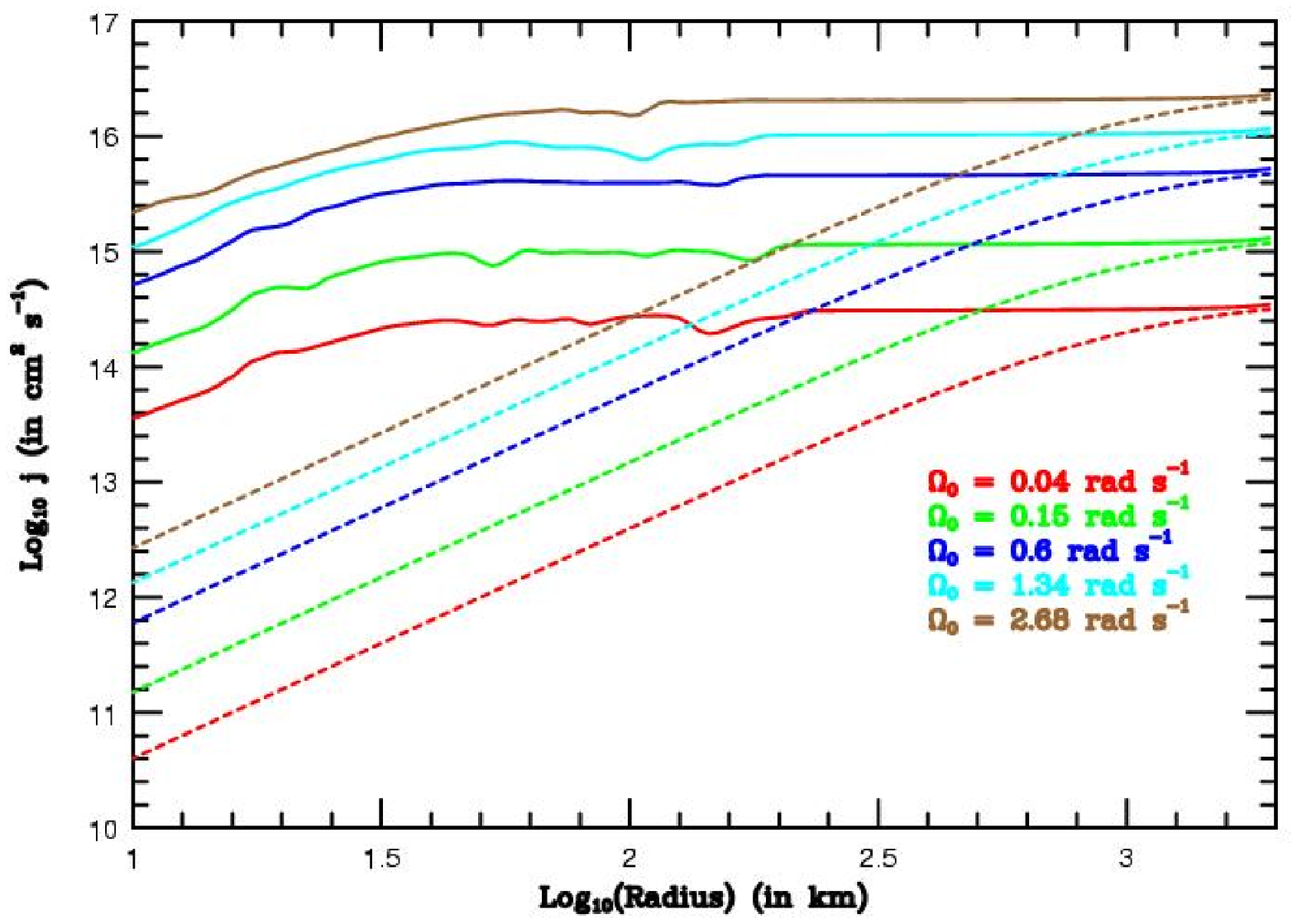}}
\caption{Equatorial angular velocities, $\Omega$, (top panel) and
         equatorial specific angular momenta, $j$ (bottom panel), as
         a function of equatorial radius for $\Omega_0$s of 2.68, 1.34, 0.6,
         0.15, and 0.04 rad s$^-1$ at 175~ms after bounce (solid lines)
         and initially (dashed lines). 
         None of our models has a ``final" rotational
         period less than $\sim$2 milliseconds.}
\label{fig:1D.angular_velocity}
\end{figure}

Figure \ref{fig:density.models}, which shows the corresponding mass density maps and velocity
fields for the different rotation laws represented by Models 
B, C, D, and F (non-rotating) at 175 milliseconds after bounce,
is more relevant to the question at hand.  From the clearly weak 
dependence of the degree of oblateness of the color contours (i.e., red, purple)
with initial rotational parameter $\Omega_0$, we see that only the 
fastest of the models in this model set  ($\Omega_0$ = 2.68 and 1.34 rad s$^{-1}$)
manifest dramatic oblateness in the isodensity contours, and this in only the inner 100 km.
Figure \ref{fig:1D.angular_velocity} portrays the corresponding ``final" (at 175 ms) angular
velocity and specific angular momentum profiles for Models A-E,F.  Since the
models depicted were generated using 2D MGFLD with multiple neutrino species
and realistic opacities, 2D hydrodynamics with rotation, and
a realistic equation of state (Lattimer \& Swesty 1991), this figure gives the most 
accurate mapping between initial rotational law and ``final" 
protoneutron star rotation profile yet calculated. 
Even with $\Omega_0$ = 0.6 rad s$^{-1}$ (initial period, P$_0$, $\sim$10 seconds),
Fig. \ref{fig:density.models} shows no dramatic rotational flattening,
though at 175 ms the equator-to-pole density ratio at a fixed radius
of 90 km is $\sim$1.5, at a fixed radius of 50 km is $\sim$1.4, 
and at a fixed radius of 30 km is $\sim$1.2.  The corresponding ratio 
at 30 km and 175 ms for $\Omega_0$ = 1.34 rad s$^{-1}$ is $\sim$2.0.
We conclude that when one uses more sophisticated 2D multi-group
neutrino transport instead of leakage or gray schemes, and consistently 
incorporates rotation into the dynamics of collapse and shock formation, 
models with ``final" core spin periods greater than $\sim$10 milliseconds
($\Omega_0 \le 0.6$ rad s$^{-1}$) show only modest rotational distortions in the inner core.  This does not
mean that rotation does not have dynamical effects.  Rather, it means merely that only
rapid rotation (as quantified here) can induce significant matter oblateness
near the neutrino decoupling surfaces.  Note that Models A and B have initial
spin rates at the upper end of those generated by Hirschi, Meynet, \& Maeder (2004).

Figure \ref{fig:density.models} also demonstrates that the number
of convective rolls and plumes increases with spin rate.  At $\Omega_0 = 0$,
only large-scale convective modes predominate.  However, for $\Omega_0 = 1.34$ rad s$^{-1}$,
the number of rolls hovers around five at 175 ms, as does the number
of downwelling plumes.  The barrel-shaped structures rotating on cylinders
that are created after bounce are broken up by these rolls in a classic
pattern.  Note, however, that for the most rapidly rotating Model A 
the number of rolls is actually smaller than for Model B.  
Consistent with the Solberg-H{\char'034}iland stability condition, convection
in rapid Model A is partially stablized in the gain region.  Also, the polar heating rate
for Model A is sufficiently high relative to its equatorial heating rate that the polar entropy 
is always much higher than that off the poles.  
Furthermore, buoyancy behind the shock keeps the hot bubble 
generated by neutrino heating near the poles confined there.

\begin{figure}
\figurenum{4}
\centerline{ \includegraphics[width=18.cm]{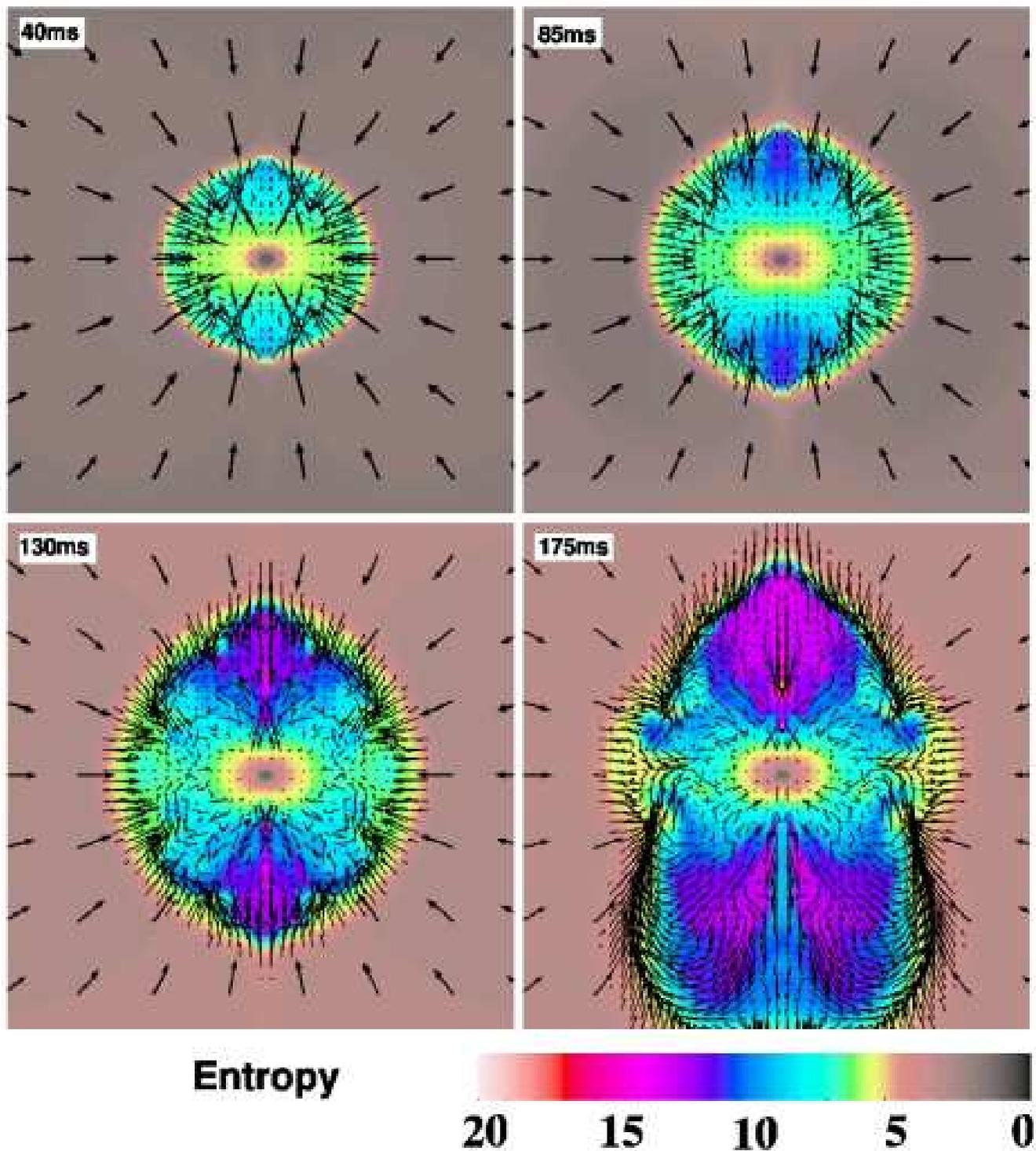}  }
\caption{Color map of the entropy (per baryon per Boltzmann's constant),
         with the r-z velocities (arrows) superposed, of the fast rotating model A ($\Omega =
         2.68$ s$^{-1}$).  Shown is the inner 600~km on a side at times
         40~ms (upper left panel), 85~ms (upper right panel), 130~ms
         (lower left panel), and 175~ms (lower right panel) after
         bounce. The entropy in the polar direction is about a factor of
         two higher than in the equatorial regions at the same
         radius. However, a high-entropy wedge is clearly widening
         as time proceeds. Note the pronounced oblateness of the
         low entropy core.  (Compare with the entropy maps of the
         models with lower rotation rates in Fig.~\ref{fig:entropy.models}.)}
\label{fig:entropy.omega.2.68.times}
\end{figure}

Figure \ref{fig:entropy.omega.2.68.times} depicts the entropy distribution of Model A
as a function of time.  We see that early in the model's post-bounce evolution
the entropy along the poles is larger (purple) than along the equator.  This is
a manifestation of the larger heating rate along the poles caused by the
rotation-induced asymmetry of the neutrino flux and heating rates (\S\ref{fluxaniso}).  For Model A
after $\sim$85 ms, an angular region that extends approximately 
45$^{\circ}$ from the pole clearly has higher entropies.  In these simulations,
there is a slight axis anomaly due to resolution and finite-difference inaccuracies
when we use cylindrical coordinates. However, this region extends only a few degrees on either side
of the pole and does not explain the much-wider-angle 
high-entropy caps. Note that in the VULCAN/2D ALE scheme,
during infall the specific angular momentum is advected numerically 
along the Eulerian grid to better than $\sim$1 percent,
but in the inner core ($\le$0.3 M$_{\odot}$) the advection error can reach $\sim$10\%. 
However, by numerical construction we conserve total angular momentum exactly.

\begin{figure}
\figurenum{5}
\centerline{ \includegraphics[width=18.cm]{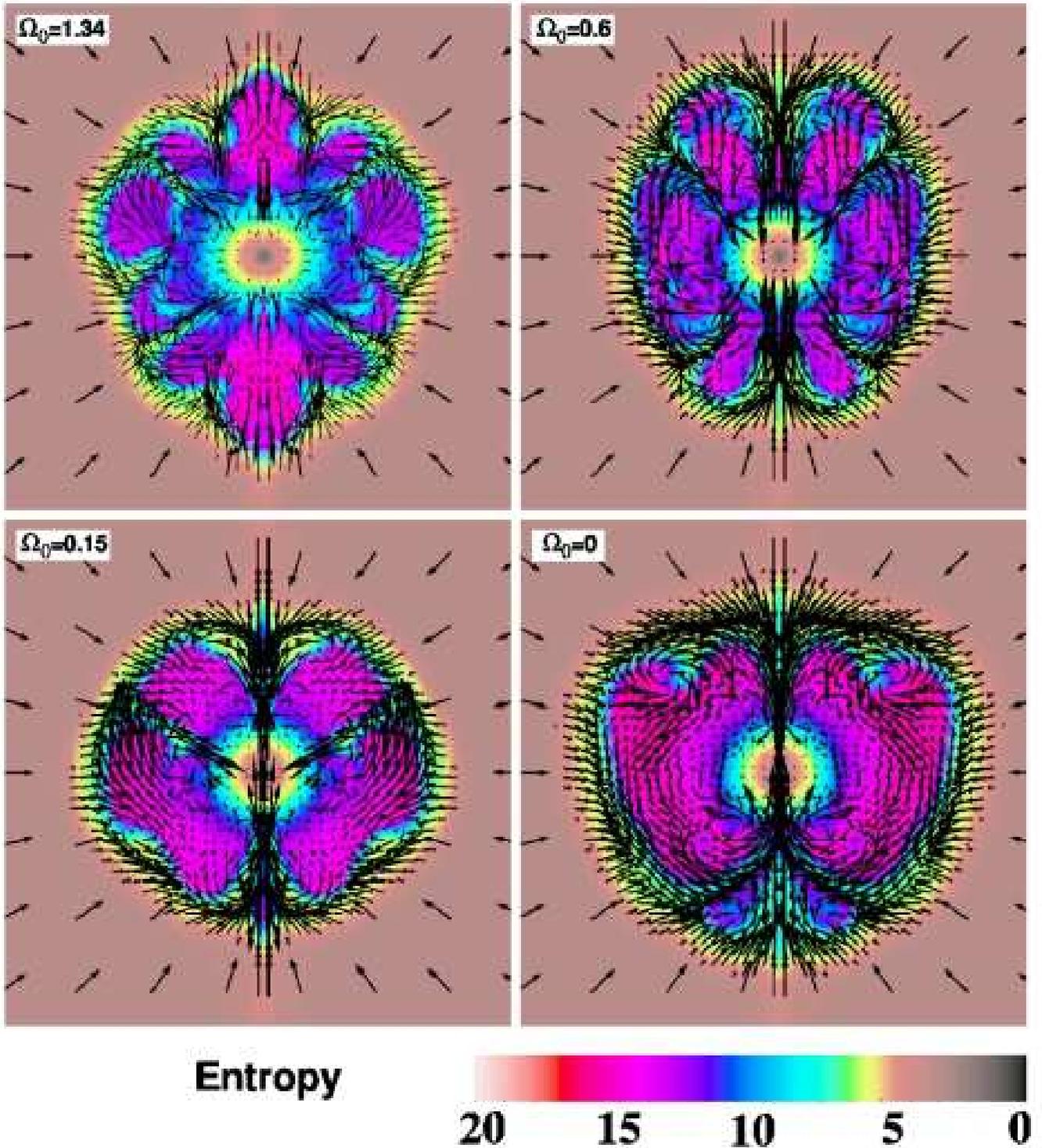} }
\caption{Same as Fig. \ref{fig:entropy.omega.2.68.times}, but 
         for Models~B ($\Omega1.34 = 2.68$ rad s$^{-1}$, upper left
         panel), C ($\Omega = 0.6$ rad s$^{-1}$, upper right panel), D
         ($\Omega = 0.15$ rad s$^{-1}$, lower left panel), and F
         (non-rotating, lower right panel) at 175 ms after bounce. The
         inner 600~km on a side is shown. Comparing with the rapidly 
         rotating Model A (with $\Omega = 2.68$ rad s$^{-1}$, shown in
         Fig.~\ref{fig:entropy.omega.2.68.times}), the entropy behind the shock is much more uniformly
         distributed; the clear contrast between equator and pole
         is absent. In the slowly rotating models, only cold convectional 
         downflows interrupt the otherwise uniform high-entropy regions.}
\label{fig:entropy.models}
\end{figure}

However, as Fig. \ref{fig:entropy.omega.2.68.times} indicates, with time the region
of high entropy spreads in angle from the poles. For Model A, the top-bottom
asymmetry emerges before the higher-entropy material has spread over the full 180$^{\circ}$
of the simulation behind the shock. But, for rotating Models B-E, though the average radius
of the shock always increases after about 100 ms after bounce, the high-entropy
rolls have spread completely around the sphere before a significant top-bottom,
dipolar asymmetry in the matter or any further dynamical effects are manifest.
Figure \ref{fig:entropy.models} portrays the entropy maps at 175 ms after bounce for 
Models B, C, D, and F (non-rotating).  Though in these models there is a slightly
greater heating rate near the poles, the material there 
with higher entropy (and, hence, higher buoyancy) spreads in convective rolls
away from the poles to larger angles.  These ``bubbles" are still confined in these 
models by the shock, before any hint of a dynamical transition is seen.  The entropy
distribution for these slower models is therefore more mixed than for Model A
with its higher spin rate and partially stabilized 
convection (see Fig. \ref{fig:entropy.omega.2.68.times}).

\section{Neutrino Flux and Heating Anisotropies due to Rotation}
\label{fluxaniso}

We now turn to a discussion of the degree of 
angular anisotropy in the neutrino field induced by
core rotation.  To demonstrate this physics it is useful to focus
on the neutrino flux, local neutrino energy density, and total net gain 
(net heating rate).  The flux and energy density are functions of energy
group and species, and all quantities are functions of time and Model.  
Rather than present all these quantities for every group, 
every timestep, every species, and every Model we have picked a few slices
in this large space to communicate the basic results.   

\begin{figure}
\figurenum{6}
\centerline{ \includegraphics[width=18.cm]{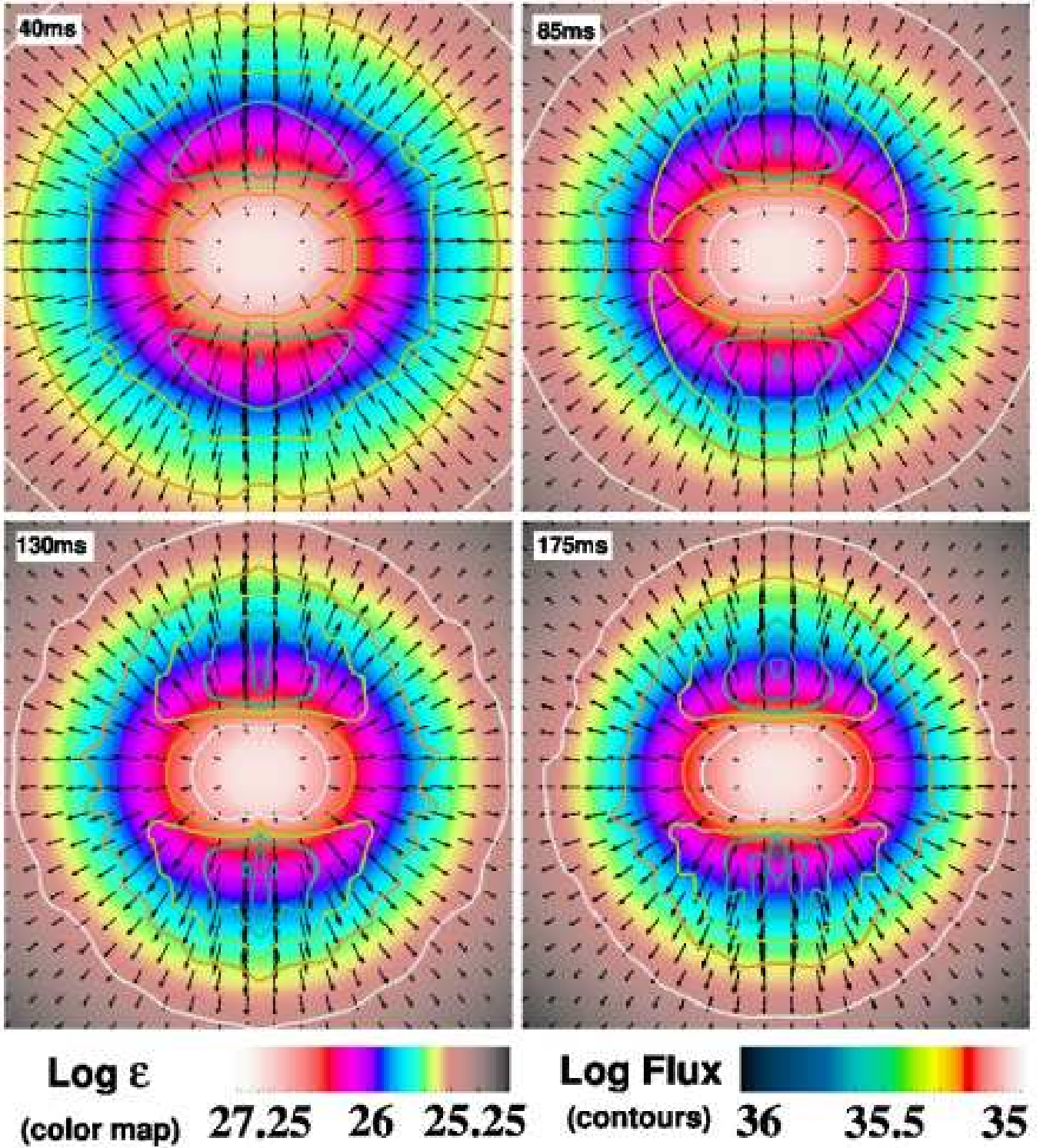} }
\caption{Time-evolution of both the distribution of the spectral energy density, $\varepsilon$, 
         of $\nu_e$-neutrinos at 6.9~MeV (color map, in
         erg cm$^{-3}$ MeV$^{-1}$) and the corresponding fluxes (contours and
         vectors, in erg cm$^{-2}$ s$^{-1}$ MeV$^{-1}$) for the fast
         rotating Model~A ($\Omega = 2.68$ rad s$^{-1}$) at times
         40~ms (upper left panel), 85~ms (upper right panel), 130~ms
         (lower left panel), and 175~ms (lower right panel) after
         bounce. The inner 240~km on a side is shown. Even for this our
         fastest rotating model, the oblateness of the energy density
         contours in the inner region is modest. In the outer regions,
         the energy density becomes slightly prolate, partially as a consequence
         of the oblateness of the neutrinospheres. Note that the equator-to-pole
         asymmetry of the flux is only moderate (at most a factor of
         two).}
\label{fig:ed.flux.e-6.9.omega.2.68}
\end{figure}

\begin{figure}
\figurenum{7}
\centerline{ \includegraphics[width=18.cm]{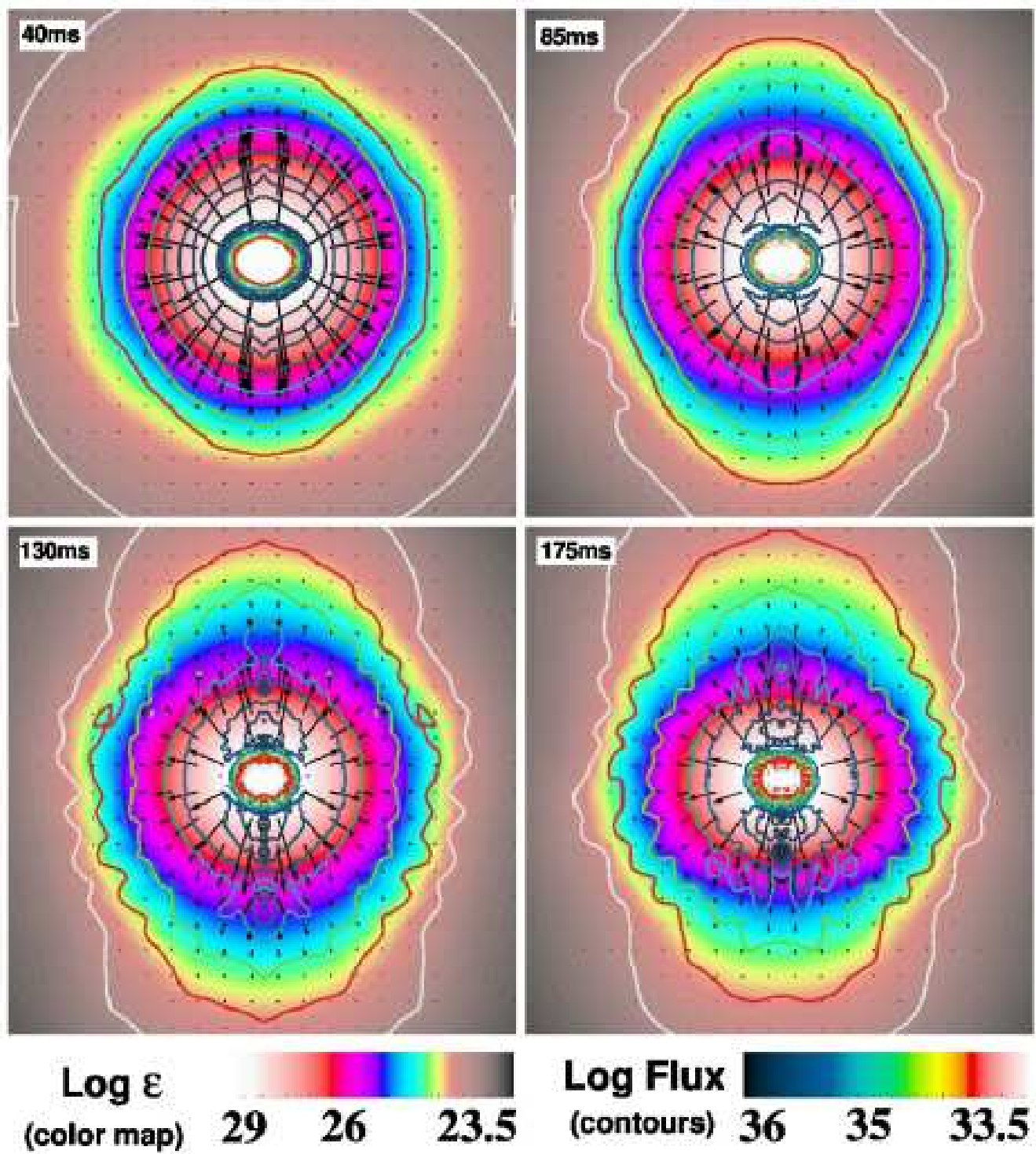} }
\caption{Same as Fig. \ref{fig:ed.flux.e-6.9.omega.2.68}, but
         for 21-MeV $\nu_e$ neutrinos.  The inner  
         240~km on a side is shown.  Due to the higher neutrino-matter
         cross section at 21 MeV, the 21-MeV neutrinosphere is at a larger radius than 
         the 6.9-MeV neutrinosphere.
         Also, the equator-to-pole flux asymmetry is slightly higher than that seen 
         in Fig.~\ref{fig:ed.flux.e-6.9.omega.2.68}, but is still modest. Furthermore, the
         contrast between the oblateness of the corresponding spectral energy density contours in
         the inner region and its prolateness further out 
         is more pronounced than for the lower
         energetic neutrinos. Both, energy density and flux decrease with
         radius much faster than for the lower energetic neutrinos.}
\label{fig:ed.flux.e-21.omega.2.68}
\end{figure}

Figure \ref{fig:ed.flux.e-6.9.omega.2.68} depicts the evolution of
the flux (in erg cm$^{-2}$ s$^{-1}$ MeV$^{-1}$; vectors and isocontours) 
and neutrino energy density (in erg cm$^{-3}$ MeV$^{-1}$)
for the $\nu_e$ neutrinos at 6.9 MeV and for high-spin Model A ($\Omega_0 = 2.68$ rad s$^{-1}$).   
The shape of the contours and the relative length of the vectors
indicate the anisotropy of the flux at 6.9 MeV.
Figure \ref{fig:ed.flux.e-21.omega.2.68} shows the
same quantities, but for a $\nu_e$ neutrino energy of
21 MeV.  The higher neutrino-matter cross section of a 21 MeV neutrino 
puts its decoupling neutrinosphere further out in radius (at $R_{\nu}$), at different densities and spin rates.  

\begin{figure}
\figurenum{8}
\centerline{ \includegraphics[width=18.cm]{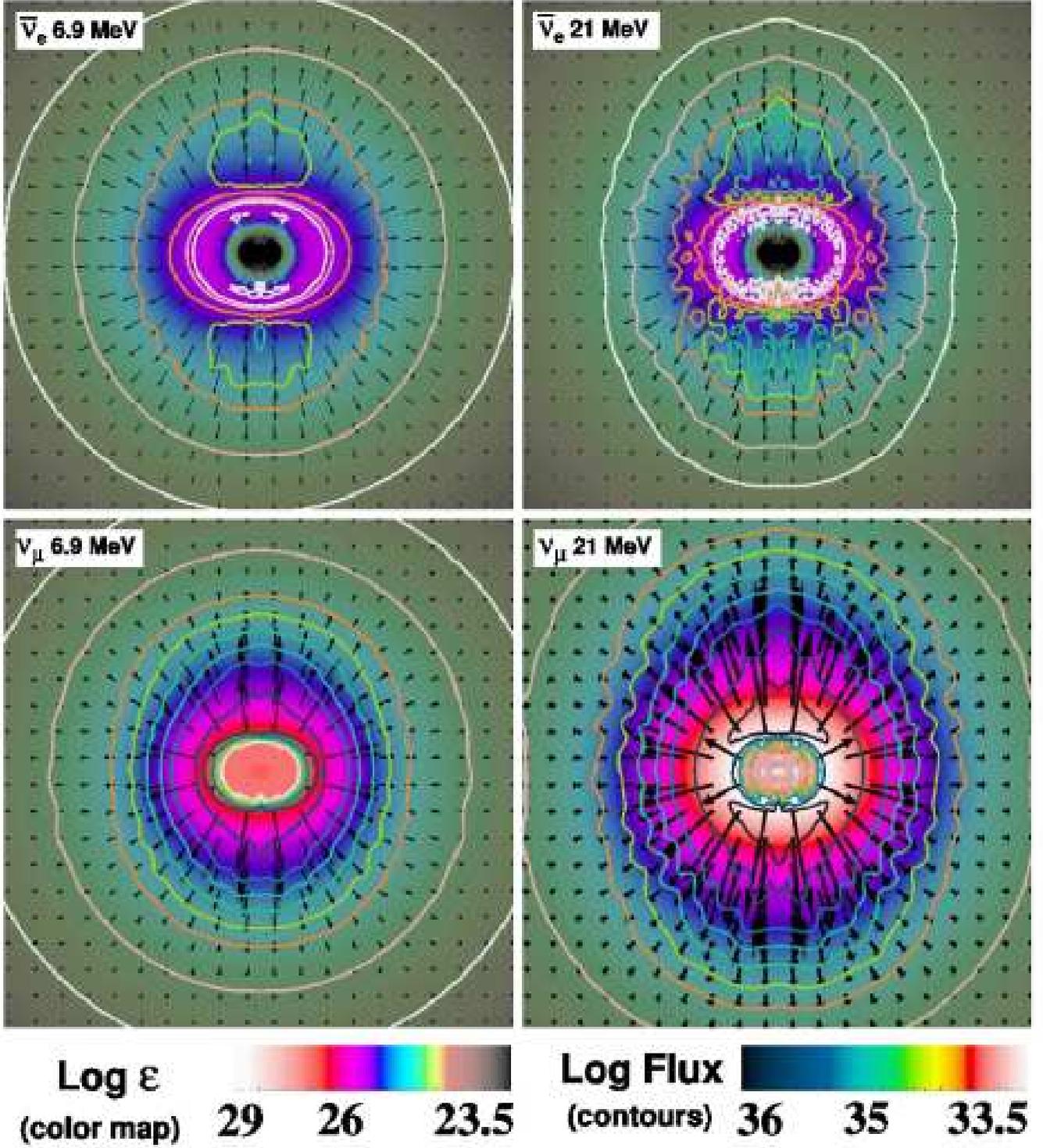} }
\caption{Snapshots at 175~ms after bounce showing for Model~A ($\Omega
         = 2.68$ rad s$^{-1}$) the spectral energy density,
         $\varepsilon$, (color map, in erg cm$^{-3}$ MeV$^{-1}$) and flux
         (contours and vectors, in erg cm$^{-2}$ s$^{-1}$ MeV$^{-1}$) of
         $\overline{\nu}_e$s at energies of 6.9~MeV (upper left panel) and 21~MeV
         (upper right panel) and of $\nu_{\mu}$s (lower panels).
         The scale for the $\overline{\nu}_e$ fluxes 
         is 2 times that for the 
         $\nu_{\mu}$ fluxes. The equator-to-pole asymmetry is slightly smaller
         then for the $\nu_e$-neutrinos.}
\label{fig:ed.flux.ep_mu.6.9_21.o2.68}
\end{figure}

The flatness of the color maps (energy density) on Fig. \ref{fig:ed.flux.e-6.9.omega.2.68}
show that the radiation field is indeed oblate, but by 175 ms has
only a $\sim$2:1 axis ratio.  That ratio gradually increases with time
as the core slowly spins up at a rate ($d\log(\Omega)/dt$) of roughly 5-10\% per 100 milliseconds.
The oblateness of the color contours is roughly consistent with Von Zeipel's
theorem for rotating stars, which states that iso-\teff surfaces 
follow equipotential surfaces.  However, further out in radius the color
contours become slightly prolate.  This is particularly clear in Fig.
\ref{fig:ed.flux.e-21.omega.2.68}.  The transition from oblate to prolate
is generic and is a consequence of the fact that at larger distances the
angle subtended at the poles by the oblate, though diffuse,  neutrinospheres is
larger than the corresponding angle at the equator. However, the pole-to-equator asymmetry 
is not as large as a naive calculation would imply.  The neutrinosphere radius ($R_{\nu}$), if such 
can be defined in 2D, is a function of neutrino energy. Moreover, for a given neutrino 
energy $\Delta R_{\nu}/R_{\nu}$ is large.  The fact that the neutrinospheres
for all species and energy groups are not sharp in radius and that there
is significant neutrino emission even exterior to a ``$\tau = 2/3$" surface
mutes the magnitude of the flux and energy density anisotropies. This is
an important effect that can be determined only using multi-D, multi-group transport. 
As the shape of the flux contours and the ratio of the vector 
lengths seen in Figs. \ref{fig:ed.flux.e-6.9.omega.2.68}
and \ref{fig:ed.flux.e-21.omega.2.68} imply, the pole-to-equator 
flux ratio at 6.9 MeV is at most a factor of $\sim$2.  At 21 MeV and a radius
of 90 km, the corresponding ratio for $\nu_e$ neutrinos is $\sim$3, 
for $\bar{\nu}_e$ neutrinos is $\sim$2.5, and for $\nu_{\mu}$ 
neutrinos is $\sim$2 (see Fig. \ref{fig:ed.flux.ep_mu.6.9_21.o2.68}).  
There is a slight tendency for the pole-to-equator flux ratio to be larger for
larger neutrino energies, but this ratio is a function of time and a strong function of radius. 

\begin{figure}
\figurenum{9}
\centerline{ \includegraphics[width=18.cm]{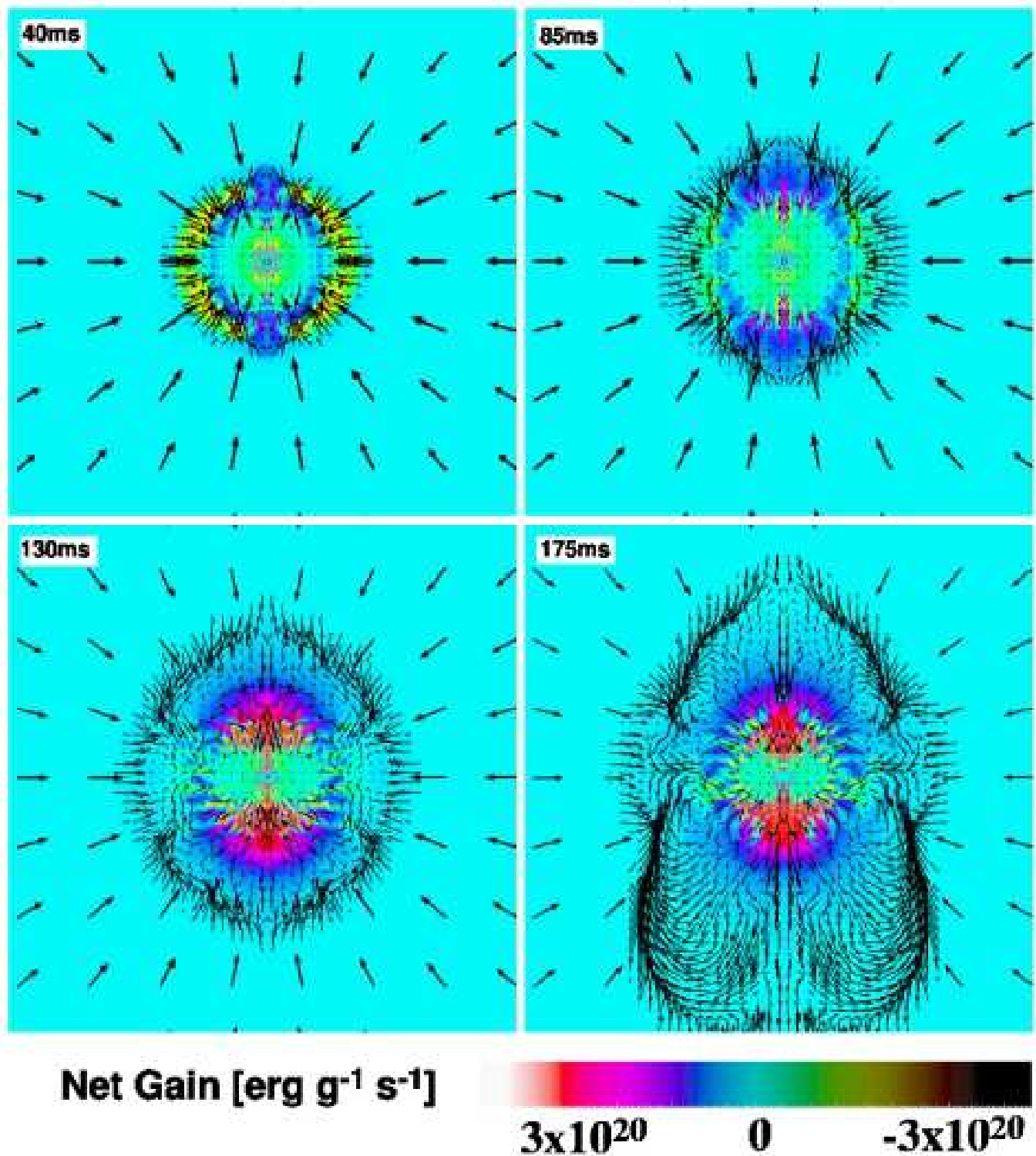}}
\caption{Integrated net energy gain (in erg g$^{-1}$ s$^{-1}$)
         due to neutrino heating for Model~A ($\Omega_0 =
         2.68$ rad s$^{-1}$) at times 40~ms (upper left panel), 85~ms
         (upper right panel), 130~ms (lower left panel), and 175~ms
         (lower right panel) after bounce.  The inner region of 600~km
         on a side is shown. The heating is much more pronounced along
         the rotation axis than at lower latitudes and manifests in fact
         the strongest equator-to-pole asymmetry of all quantities
         investigated.  Nevertheless, the net gain never varies by more than about
         a factor of a few at a given radius within the shock. Fig~\ref{fig:gain.non_rot} shows
         the clear dependence of this effect on the rotation rate.}
\label{fig:gain.omega.2.68}
\end{figure}

A color plot of the net rate of neutrino energy deposition (the net gain)
(in erg g$^{-1}$ s$^{-1}$, integrated over neutrino energy)
for Model A at different times is given in Fig. \ref{fig:gain.omega.2.68}.  In this figure,
the vectors are velocity vectors.  Red and purple represent high rates of heating,
while green and yellow represent low rates of heating or net losses.  The strong polar heating, particularly
at later times, is clearly seen and recapitulates what is seen in Fig. \ref{fig:entropy.omega.2.68.times}.
However, though the entropy profile can be smoothed, the heating profile 
can better maintain asymmetry.  Nevertheless, even for this rapidly rotating
model the pole-to-equator ratio in heating rate is no greater than 3-5 at 150 km,
an important area in the gain region.   

\begin{figure}
\figurenum{10}
\centerline{
     \includegraphics[width=18.cm]{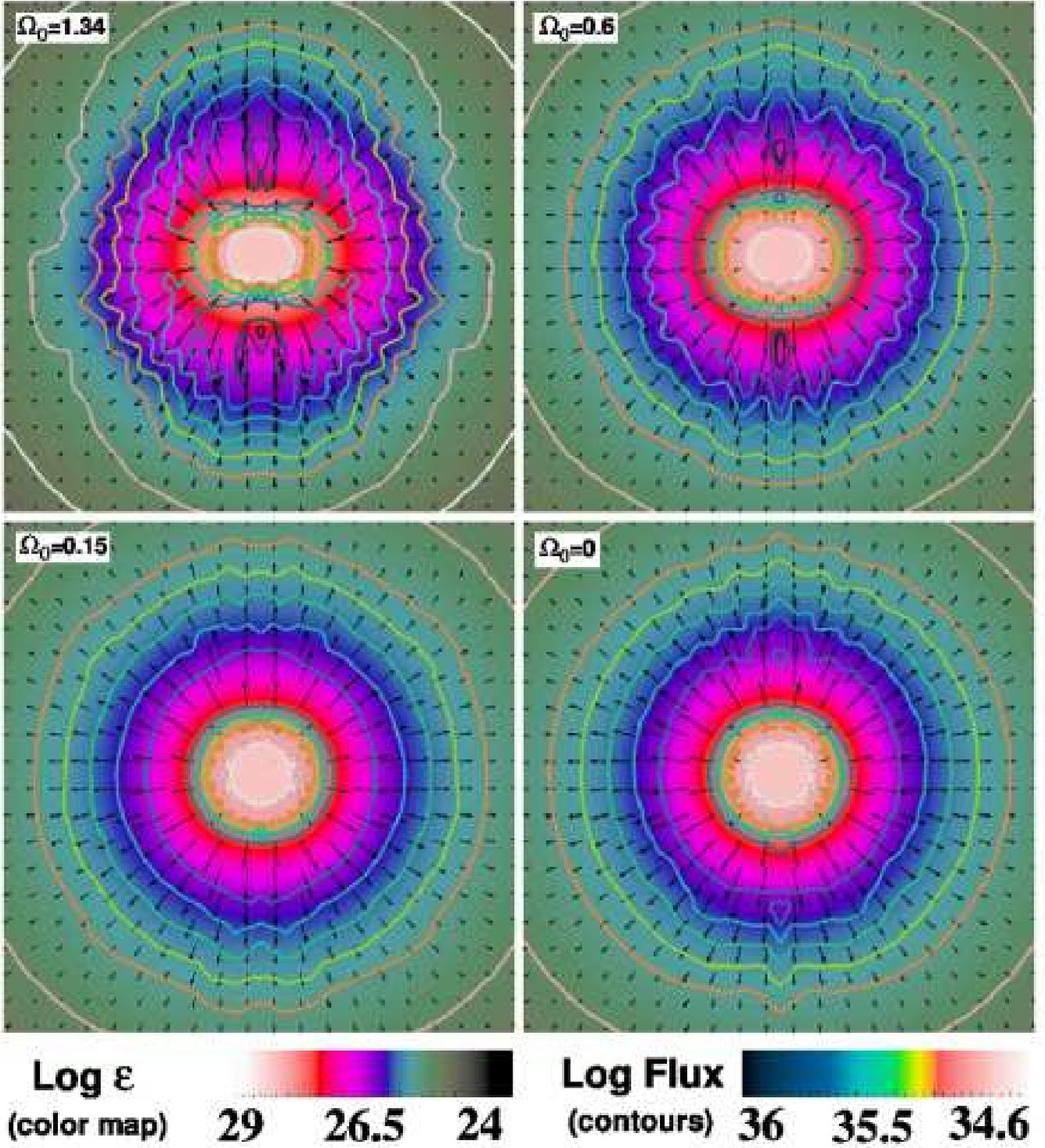}
           }
\caption{Spectral energy density, $\varepsilon$, (color map, in erg
         cm$^{-3}$ MeV$^{-1}$) and flux (contours and vectors, in erg
         cm$^{-2}$ s$^{-1}$ MeV$^{-1}$) of $\nu_{e}$s at 12~MeV for
         Models B ($\Omega_0 = 1.34$ rad s$^{-1}$, upper left panel),
         C ($\Omega = 0.6$ rad s$^{-1}$, upper right panel), D
         ($\Omega = 0.15$ rad s$^{-1}$, lower left panel), and F
         (non-rotating, lower right panel) at 175 ms after bounce.
         The inner 240~km on a side is shown. The color maps 
         should be compared to those shown in Fig.~\ref{fig:ed.flux.ep_mu.6.9_21.o2.68} and the vector
         lengths to those shown for the $\nu_{\mu}$s in
         Fig.~\ref{fig:ed.flux.ep_mu.6.9_21.o2.68}. With increasing rotation rate, the flux is
         more and more concentrated along the rotation axis.  The energy
         density distribution is oblate in the inner core and prolate
         much further out.}
\label{fig:ed.flux.e-12.diff.omega}
\end{figure}

\begin{figure}
\figurenum{11}
\centerline{ \includegraphics[width=18.cm]{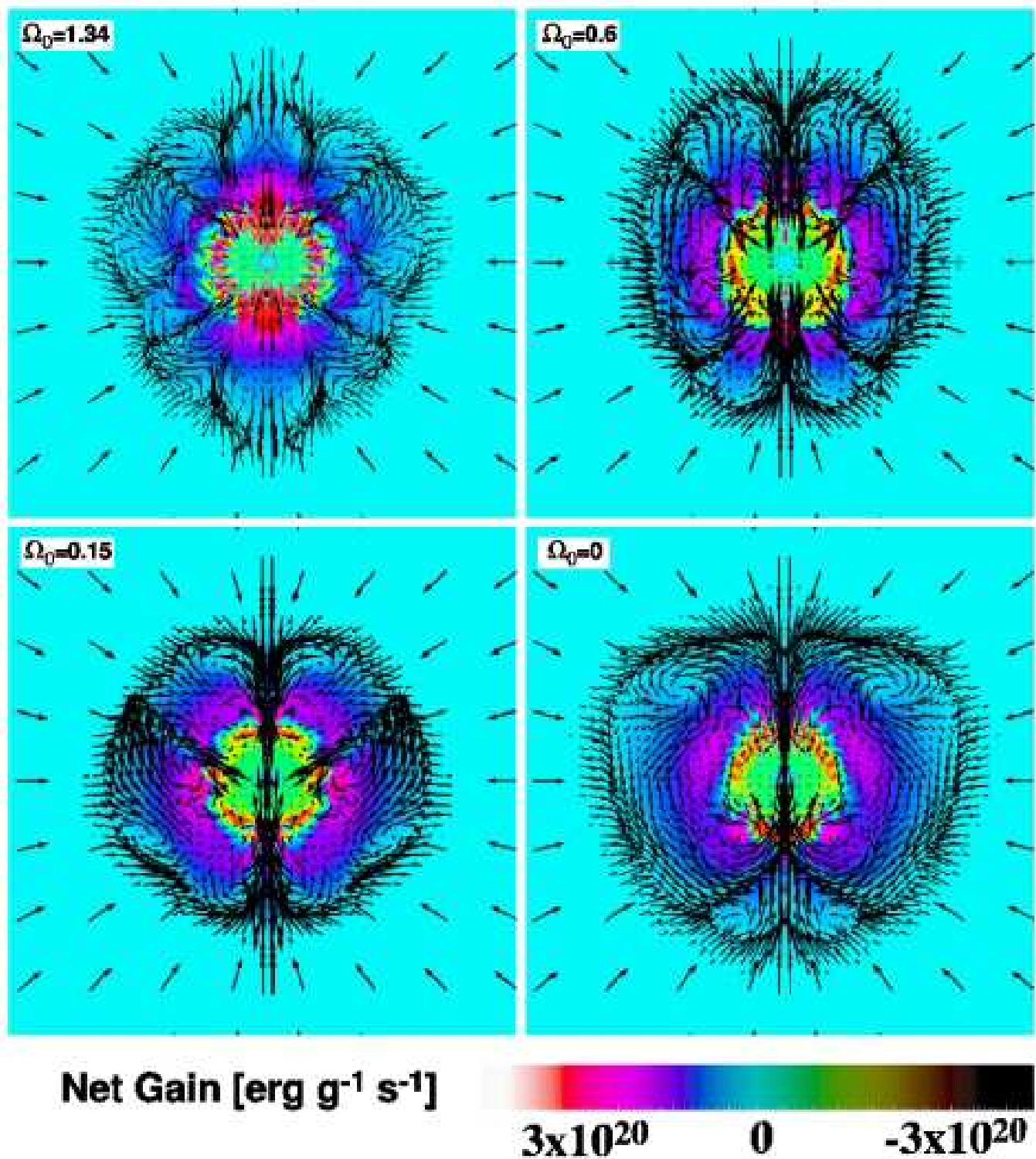} }
\caption{Integrated net energy gain (in erg g$^{-1}$ s$^{-1}$)
         due to neutrino heating for Models B ($\Omega_0
         = 1.34$ rad s$^{-1}$, upper left panel), C ($\Omega_0 = 0.6$
         rad s$^{-1}$, upper right panel), D ($\Omega_0 = 0.15$ rad
         s$^{-1}$, lower left panel), and F (non-rotating, lower right
         panel). The inner region 600~km on a side is shown. With
         increasing initial angular velocity, the heating rate is more and
         more concentrated along the poles, though this effect is only
         moderate compared with that for the rapidly rotating Model A ($\Omega =
         2.68$ rad s$^{-1}$) shown in Fig.~\ref{fig:gain.omega.2.68}.}
\label{fig:gain.non_rot}
\end{figure}

The corresponding color maps, contours, and vectors depicting the $\nu_e$ flux and 
energy density distributions (this time at 12 MeV), and the net gain distributions
for other models (B, C, D, F [non-rotating]) at 175 ms after bounce are given in 
Figs. \ref{fig:ed.flux.e-12.diff.omega} and \ref{fig:gain.non_rot}, respectively.
There is a slight flux anisotropy for the $\Omega_0 = 1.34$ rad s$^{-1}$ and $\Omega_0 = 0.6$ rad s$^{-1}$ 
models.  However, even at 1.34 rad s$^{-1}$, at 90 km the 
pole-to-equator $\nu_e$ flux ratios at 12 and 21 MeV are only $\sim$1.2.  
At 21 MeV and for $\nu_{\mu}$ neutrinos, this ratio is at most $\sim$1.4.  Even 
for the $\Omega_0 = 1.34$ rad s$^{-1}$ model, with a post-bounce spin period of 6-10 ms, the 
pole-to-equator heating rate anisotropy at 150 km 
hovers between 1.0 and 2.5. We conclude that while there
is a greater heating rate at the poles for rotating models, the
rotation rate required for a significant effect is large.  Only our Model A
shows a significant effect, though the magnitude of 
this effect as measured by the heating rate asymmetry 
and pole-to-equator flux ratio are not as large as estimated
in the previous literature.

\section{Conclusions}
\label{conclusion}

Using the 2D multi-group, flux-limited diffusion version of the code VULCAN/2D (Livne et al. 2004),
we have calculated the collapse, bounce, shock formation, and early post-bounce evolutionary
phases of a core-collapse supernova for a variety of initial rotation rates.  This is the first
series of such multi-group calculations undertaken in supernova theory with fully multi-D tools.  
We find that rotation does indeed generate pole-to-equator anisotropies in the neutrino
radiation fields and fluxes, but that the magnitude of the asymmetry is not as large as
previously estimated.  The finite width of the neutrino decoupling surfaces and the
broad distribution of neutrino sources above the $\tau=2/3$ surface mute the angular contrast.
We have explored the angular dependence of the neutrino fields 
as a function of neutrino species, neutrino energy, and initial rotation
rate.  Only for our most rapidly rotating model (with $\Omega_0 = 2.68$ rad s$^{-1}$) do
we start to see qualitatively different hydrodynamics, but for the lower rates consistent
with the pre-collapse rotational profiles derived in the literature the anisotropies
are rather more tame than anticipated.  In addition, we have not been able to reproduce the suggestion
of Shimizu \etal (2001) and Madokoro, Shimizu, \& Motizuki (2004) that even a 
a few to a few tens of percent neutrino flux anisotropy can have a demonstrable effect on the hydrodynamics.
This does not mean that rotation can not play
a key role in collapse and supernova dynamics (Burrows, Ott, \& Meakin 2003;
Burrows \etal 2004; Yamasaki \& Yamada 2004).  The decrease in the effective gravity due to the centripetal 
effect can be quite important.  Rather, it means that when a realistic mapping between initial and final 
rotational profiles and 2D multi-group radiation-hydrodynamics are incorporated into collapse simulations
the anisotropy of the radiation fields may be only a secondary, not a pivotal factor, 
in the supernova mechanism.  Moreover, we find that the radiation field is always more smooth
and symmetric than the matter distribution, with its plumes and convective eddies.
The radiation field at a point is an integral over many sources from the different contributing directions.
As such, it does not vary as much as the matter on small spatial scales and has very little power
at high spatial frequencies.  The larger spatial and temporal variations 
in the neutrino flux are seen for the higher energy groups.

\acknowledgments

We acknowledge discussions with and help from 
Jeremiah Murphy, Casey Meakin, Salim Hariri, 
Marvin Landis, Thomas Janka, and Stan Woosley.  We thank the 
Institute for Nuclear Theory (INT) of the University of Washington
for their kind hospitality in the summer of 2004, during 
which some of this paper was incubated. 
Importantly, we acknowledge support for this work
from the Scientific Discovery through Advanced Computing 
(SciDAC) program of the DOE, grant number DE-FC02-01ER41184. R.W. thanks 
the Institute of Astronomy of ETH Zurich for providing
part-time office space, E.L. thanks the Israel Science Foundation
for support under grant \# 805/04, and C.D.O. thanks the Albert-Einstein-Institut
for providing CPU time on their Peyote Linux cluster. The AEI publication number
is AEI-2005-001.
Finally, we thank Jeff Fookson and Neal Lauver of the Steward Computer Support Group
for their invaluable help with the local Beowulf cluster and acknowledge
the use of the NERSC/LBNL/seaborg and ORNL/CCS/cheetah machines.
Movies and still frames associated with this work can be obtained
upon request.

\begin{deluxetable}{ccccccc}
\tablecaption{Initial Model Parameters\label{table:models}}
\tablehead{
\colhead{Model Name}&
\colhead{$\Omega_0$}&
\colhead{$A$}&
\colhead{Central $P_0$}&
\colhead{$P_0$ at 2000 km}&
\colhead{$T/|W|_i$}&
\colhead{$T/|W|_f$}\\
\colhead{}&
\colhead{(rad s$^{-1}$)}&
\colhead{(km)}&
\colhead{(s)}&
\colhead{(s)}&
\colhead{(\%)}&
\colhead{(\%)}
}
\startdata
A&2.68&1000&2.34&11.70&0.29& 6.50\\
B&1.34&1000&4.69&23.45&0.075& 2.25\\
C&0.60&1000&10.47&52.36&0.015& 0.50\\
D&0.15&1000&41.89&209.44&9.4 $\times$ 10$^{-4}$&$3.6\times 10^{-2}$\\
E&0.04&1000&157.1&785.40&6.7 $\times$ 10$^{-5}$&$2.7\times 10^{-3}$\\
F&0.0& - & - & - & - & - \\
\enddata
\tablecomments{
$\Omega_0$ and $A$ are the parameters used in eq. (\ref{eq:rotlaw})
to define the initial rotational profiles and $P_0$ is the initial period. 
Here, $T/|W|_i$ is for the initial configuration and $T/|W|_f$ 
is for the final configuration.  See text for discussion.}
\end{deluxetable}

\clearpage


\clearpage

\clearpage

\clearpage

\clearpage

\clearpage

\clearpage

\clearpage

\clearpage

\clearpage

\clearpage


\begin{thebibliography}{}



\bibitem[Bethe \& Wilson 1985]{bethe}
Bethe, H. \& Wilson, J.~R.~1985, \apj, 295, 14

\bibitem[Bruenn 1985]{BR1} Bruenn, S.W. 1985, \apjs, 58, 771

\bibitem[Buras et al. 2003]{buras2003} Buras, R., Rampp, M., Janka, H.-Th., \& Kifonidis, K. 2003,
\prl, 90, 241101

\bibitem[Burrows \etal  2000]{B2} Burrows, A., Young, T., Pinto, P., Eastman,
        R. \& Thompson, T. 2000, \apj, 539, 865

\bibitem[Burrows,~Hayes,~\&~Fryxell 1995]{bhf}
Burrows, A., Hayes, J., \& Fryxell, B.A.~1995, \apj, 450, 830

\bibitem[Burrows,~Ott,~\&~Meakin 2003]{bom} Burrows, A., Ott, C.D., \& Meakin, C. 2003,
to be published in the proceedings of ``3-D Signatures in Stellar Explosions:
A Workshop honoring J. Craig Wheeler's 60th birthday," held June 10-13, 2003, Austin, Texas, USA

\bibitem[Burrows et al. 2004]{bur2004} Burrows, A., Walder, R., Ott, C.D.,
and Livne, E. 2004, ``Rotating Core Collapse and Bipolar Supernova Explosions,"
to be published in the proceedings of the international conference entitled ``The Fate of the
Most Massive Stars," held May 23-28, 2004, at Jackson Lake Lodge in the
Grand Teton National Park, Wyoming, USA, ed. Roberta Humphreys (ASP Conf. Series),
astro-ph/0409035.

\bibitem[Burrows et al. 2005]{bur2005} Burrows, A. et al. 2005, in preparation

\bibitem[Fryer \& Heger 2000]{fryer2000} Fryer, C.L. \& Heger, A. 2000, \apj, 541, 1033

\bibitem[Fryer \& Warren 2002]{fryer2002} Fryer, C.L. \& Warren, M. 2002, \apj, 574, L65

\bibitem[Fryer \& Warren 2004]{fryer2004} Fryer, C.L. \& Warren, M. 2004, \apj, 601, 391

\bibitem[Heger,~Langer,~\&~Woosley 2000]{heger00}
{Heger}, A., {Langer}, N., and {Woosley}, S.E. 2000, \apj, 528, 368

\bibitem[Heger,~Woosley,~\&~Langer 2003]{langer}
Heger, A., Woosley, S.E., \& Langer, N. 2003, in ``A Massive Star Odyssey:
From Main Sequence to Supernova," Proceedings of IAU Symposium \#212,
held 24-28 June 2001 in Lanzarote, Canary Islands, Spain.
Edited by Karel van der Hucht, Artemio Herrero, and C\'{e}sar Esteban (San Francisco:
Astronomical Society of the Pacific), p.357

\bibitem[Heger,~Woosley,~\&~Spruit 2004]{heger04} Heger, A., Woosley, S.E., \& Spruit, H. 2004, astro-ph/0409422

\bibitem[Herant et al. 1994]{herant}
Herant, M., Benz, W., Hix, W.R., Fryer, C.L., \& Colgate, S.A. 1994, \apj, 435, 339

\bibitem[Hirschi,~Meynet,~\&~Maeder 2004]{hirschi} Hirschi, R., Meynet, G., \& Maeder, A. 2004,
submitted to \aa\ (astro-ph/0406552)

\bibitem[Hwang et al. 2004]{hwang04} Hwang, U. et al. 2004, \apj, in press.

\bibitem[Janka \& M\"onchmeyer 1989a]{1monch89} Janka, H.-T. \& M\"onchmeyer 1989a, Astron. \& Astrophys., 209, L5

\bibitem[Janka \& M\"onchmeyer 1989b]{monch89} Janka, H.-T. \& M\"onchmeyer 1989b, Astron. \& Astrophys., 226, 69

\bibitem[Janka,~Buras,~\&~Rampp 2003]{janka2003} Janka, H.-T., Buras, R., \& Rampp, M. 2003, Nucl. Phys. A, 718, 269

\bibitem[Janka et al. 2004]{janka04} Janka, H.-T., Scheck, L., Kifonidis, K., M\"uller, E., \& Plewa, T. 2004,
astro-ph/0408439 

\bibitem[Kotake et al. 2003]{kotake} Kotake, K., Yamada, S., \& Sato, K. 2003, \apj, 595, 304

\bibitem[Lattimer \& Swesty 1991]{Lat1} Lattimer, J.M. \& Swesty, F.D., 1991 Nucl. Phys. A, 535,331

\bibitem[Liebend\"{o}rfer et al. 2001]{lieben2001}
Liebend\"{o}rfer, M., Mezzacappa, A., Thielemann, F.-K., Messer,
O. E. B., Hix, W.~R., \& Bruenn, S.W.~2001, \prd, 63, 103004

\bibitem[Livne et al. 2004]{livne04}Livne, E., Burrows, A., Walder, R.,
Thompson, T.A., and Lichtenstadt, I. 2004, \apj, 609, 277

\bibitem[Madokoro,~Shimizu,~\&~Motizuki 2004]{madokoro} Madokoro, H.,
Shimizu, T., \& Motizuki, Y. 2004, astro-ph/0312624

\bibitem[Meynet,~Hirschi,~\&~Maeder 2004]{meynet} Meynet, G., Hirschi, \& Maeder, A. 2004, astro-ph/0409508

\bibitem[Mezzacappa et al. 2001]{mezz2001}
Mezzacappa, A., Liebend\"{o}rfer, M., Messer, O.E.B.,
Hix, W.R., Thielemann, F.-K., \& Bruenn, S.W.~2001, \prl, 86, 1935

\bibitem[Ott et al. 2004]{ott} Ott, C.D., Burrows, A., Livne, E., \& Walder, R. 2004,
\apj, 600, 834

\bibitem[Rampp \& Janka 2000]{rampp2000} Rampp, M. \& Janka, H.-T. 2000, \apj, 539, L33

\bibitem[Shimizu et al. 2001]{shimizu} Shimizu, T., Ebisuzaki, T., Sato, K., \& Yamada, S. 2001, \apj, 552, 756

\bibitem[Scheck et al. 2004]{scheck}
Scheck, L., Plewa, T., Janka, H.-Th., Kifonidis, K., \& M\"uller, E. 2004, \prl, 92, 011103

\bibitem[Thompson,~Burrows,~\&~Pinto 2003]{Tod1} Thompson, T.A., Burrows, A., \& Pinto, P.A., 2003, \apj, 592, 434


\bibitem[Wang et al. 2002]{wang1}
Wang, L., et al. 2002, \apj, 579, 671 

\bibitem[Wang et al. 2003]{wang2}
Wang, L., Baade, D., H\"oflich, P., \& Wheeler, J.C. 2003, \apj, 592, 457 

\bibitem[Willingale et al. 2002]{will02} Willingale, R., Bleeker, J.A.M., van der Heyden, K.J.,
Kaastra, J.S., \& Vink, J. 2002, \aa, 381, 1039

\bibitem[Willingale et al. 2003]{will03} Willingale, R., Bleeker, J.A.M., van der Heyden, K.J.,
\& Kaastra, J.S. 2003, \aa, 398, 1021

\bibitem[Woosley \& Weaver 1995]{woosley}
Woosley, S.E. \& Weaver, T.A. 1995, \apjs, 101, 181

\bibitem[Yamasaki \& Yamada 2004]{yama} Yamasaki, T. \& Yamada, S. 2004, astro-ph/0412625


\end{thebibliography}
\end{document}